# Characterizing and Detecting Freezing of Gait using Multi-modal Physiological Signals

Ying Wang, Floris Beuving, Jorik Nonnekes, Mike X Cohen, Xi Long, Ronald M Aarts, *Fellow, IEEE*, Richard Van Wezel

*Abstract*—Freezing-of-gait a mysterious symptom of Parkinson's disease and defined as a sudden loss of ability to move forward. Common treatments of freezing episodes are currently of moderate efficacy and can likely be improved through a reliable freezing evaluation. Basic-science studies about the characterization of freezing episodes and a 24/7 evidence-support freezing detection system can contribute to the reliability of the evaluation in daily life. In this study, we analyzed multi-modal features from brain, eye, heart, motion, and gait activity from 15 participants with idiopathic Parkinson's disease and 551 freezing episodes induced by turning in place. Statistical analysis was first applied on 248 of the 551 to determine which multi-modal features were associated with freezing episodes. Features significantly associated with freezing episodes were ranked and used for the freezing detection. We found that eye-stabilization speed during turning and lower-body trembling measure significantly associated with freezing episodes and used for freezing detection. Using a leave-one-subject-out cross-validation, we obtained a sensitivity of 97% ± 3%, a specificity of 96% ± 7%, a precision of 73% ± 21%, a Matthews correlation coefficient of 0.82 ± 0.15, and an area under the Precision-Recall curve of 0.94 ± 0.05. According to the Precision-Recall curves, the proposed freezing detection method using the multi-modal features performed better than using single-modal features.

This work is part of the research program BrainWave with project number 14714, which is (partly) financed by the Netherlands Organization for Scientific Research (NWO) and a research grant under the Operational Program European Regional Development Fund (OP ERDF) of the European Union. (Corresponding author: Ying Wang.)

Y. W. is with Donders Institute for Brain, Cognition and Behaviour, Radboud University, Nijmegen, the Netherlands; Department of Electrical Engineering, Eindhoven University of Technology, Eindhoven, the Netherlands; Academic Center for Epileptology Kempenhaeghe, Heeze, the Netherlands (e-mail: imwywk@gmail.com).

F.W. was with Donders Institute for Brain, Cognition and Behaviour, Radboud University, Nijmegen, the Netherlands. (florisbeuving@gmail.com).

J.N. is with Donders Institute for Brain, Cognition and Behaviour, Department of Rehabilitation, Radboud University Medical Centre, Nijmegen, the Netherlands; Department of Rehabilitation, Sint Maartenskliniek, Nijmegen, the Netherlands (email: Jorik.Nonnekes@radboudumc.nl).

M. C. is with Donders Institute for Brain, Cognition and Behaviour, Radboud University, Nijmegen, the Netherlands (email: mikexcohen@gmail.com).

X. L. is with Department of Electrical Engineering, Eindhoven University of Technology, Eindhoven, the Netherlands; Philips Research, Eindhoven, the Netherlands (email: X.Long@tue.nl).

R.A. is with Department of Electrical Engineering, Eindhoven University of Technology, Eindhoven, the Netherlands (email: R.M.Aarts@tue.nl);

R.W. is with Donders Institute for Brain, Cognition and Behaviour, Radboud University, Nijmegen, the Netherlands; Biomedical Signals and Systems, Faculty of Electrical Engineering, Mathematics and Computer Science, University of Twente, Enschede, the Netherlands (email: R.vanWezel@donders.ru.nl).

## I. INTRODUCTION

FREEZING of gait (FOG) is a common clinical symptom observed in the moderate and advanced phase of Parkinson's disease and defined as a "brief, episodic absence or marked reduction of forward progression of the feet despite the intention to walk" [1]. The sudden absence of the ability to move forward could lead to frequent falls, and the associated physical and psychosocial consequences (e.g., a bone fracture, a head injury, and fear of falling) have a huge impact on patients' quality of life [2]. Nevertheless, common treatments for freezing episodes are of moderate efficacy, such as the resistance to dopaminergic treatment and ineffectiveness of continuous external rhythmic cues [1]–[3]. The freezing treatment can be improved through a reliable freezing evaluation. A barrier to the reliable evaluation is the unpredictable, idiosyncratic, and episodic nature of freezing episodes [4]. An online evidence-support freezing monitoring system, which detects or predicts the spontaneous freezing episodes when or before they occur, can contribute the reliability of freezing evaluation in daily life; for example, external cues could be triggered on demand to help individuals overcome freezing episodes in daily life.

Physiological features extracted from wearable sensors play a key role in the monitoring of freezing episodes. Physiological signals reflecting the function of motor, cognitive, and autonomic nervous system were found to be related with freezing episodes in previous studies [5]–[8].

1) The function of the motor system is normally measured by three-dimensional (3D) gyroscopes and/or accelerometers placed on the lower body. A well-known feature "freeze index" are commonly used to describe the trembling of individual's lower body parts during movements [5], and the value of freeze index increased during freezing episodes [9]. A previous study [9] proposed an online freezing detection system using the freezing index feature and achieved a sensitivity of 73% and a specificity of 82%.
2) Brain activity extracted from scalp electroencephalography (EEG) signals was mainly investigated by a group of researchers [6], [10], [11]. They asserted a significant increase of theta-wave band (4-8 Hz) power in EEG signals between the central and frontal electrodes [6]. Based on the spatial, spectral, and temporal features of EEG signals, offline freezing detection [10] or prediction [11] systems were developed from a manually selected EEG dataset including 400-s signals from each individual phase of freezing episodes: normal walking, transition, and freezing episodes.

3) The function of autonomic nervous system was captured by the signals of electrocardiography (ECG) and/or galvanic skin response sensors [7], [8]. Heart rate extracted from ECG signals was found to considerably increase before and during freezing episodes [7]. The galvanic skin response signals showed a significant increase before freezing episodes, and a corresponding offline system predicted freezing episodes 4 s on average before they happened with a sensitivity of 71% and a precision of 65% [8].

However, earlier research [6], [7] only explored the difference among the phases of freezing episodes (normal walking, transition, and freezing episodes). This comparison method, which does not consider the baseline movement condition (such as, turning, standing, and walking), may introduce movement artefacts caused by experiment tasks sensitive to trigger freezing episodes [12]. For example, the changes of the brain activity or the heart rate may be caused by a transition from normal walking to turning instead of a freezing episode. Moreover, the studies about online monitoring systems for daily use are scarce, and challenges in accurate freezing monitoring still exist given the movement artifacts in the physiological signals and highly heterogeneous clinical symptoms in individuals [9], [13].

A reliable FOG evaluation is still difficult, especially in daily life. More basic science and engineering research is therefore desired to improve the reliability of freezing evaluation. In our preliminary study [14], we proposed an online multi-modal freezing system to detect freezing episodes in daily life. The system used brain activity extracted from EEG signals and motion activity extracted from accelerometer signals, and we found improved freezing detection performance compared to single-modal detection systems. In this research, we took two steps further to improve the reliability of the freezing detection system through combining a basic science and a applied science study: (1) To be supported by evidence, we applied statistical tests to determine whether the following multi-modal physiological activities are significantly associated with freezing episodes—eye movements, brain, heart, motion, and gait activity—and to check whether the earlier findings about brain [6] and heart activity [7] can be duplicated when the baseline movement condition is considered. (2) We applied the findings of step (1) in the development of an online freezing detection system.

## II. METHOD

### A. Ethical approval

This cross-sectional study was ethically approved by the Dutch committee on research involving human participants [Arnhem-Nijmegen region (NL60942.091.17)], and the experiment conformed to Declaration of Helsinki, and all participants provided informed consent.

### B. Participants

We recruited 17 participants with idiopathic Parkinson's disease and experiencing daily freezing episodes. The participants were clinically examined to evaluate their clinical characteristics: the subsection III of Movement Disorders Society-Unified Parkinson's disease Rating Scale (MDS-UPDRS III; including Hoehn and Yahr stage) [15]–[17] to examine the movement performance, New freezing of Gait Questionnaire (N-FOGQ) [18] to quantify freezing of gait severity, Mini-Mental State Examination (MMSE) [19] to measure cognitive impairment, and Frontal Assessment Battery (FAB) [20] to evaluate the frontal lobe performance. We included the participants who experienced regular freezing ("very often, more than one time a day" in the freezing frequency section of N-FOGQ [18]) in the past month and were able to walk 150 m independently. The participants with comorbidities that cause severe gait impairment and severe cognitive impairments (the score of a MMSE [19] <24) were excluded.

Fifteen of the 17 participants' data were collected in the study because the other two participants were unable to walk or keep balance independently during the experiments. The age of the 15 participants ranged from 51 to 89 years. Their Parkinson's disease duration ranged from 3 to 20 years, and their clinical characteristics were from 24 to 49 of MDS-UPDRS III, from 2 to 4 of the Hoehn and Yahr stage, from 10 to 25 of the N-FOGQ, from 24 to 30 of the MMSE, and from 14 to 18 of the FAB. Fourteen participants reported in the N-FOGQ that they experienced freezing during turning in daily life, and eight participants had more than one freezing episode during turning each day.

### C. Study paradigm

Participants were assessed in the dopaminergic OFF-medication state, after 12 hours of medication withdrawal. Given that freezing episodes are most sensitive to turning conditions [21], the data were collected during two turning tasks: 180-degree alternative turning at a self-selected speed and at a rapid speed. To efficiently provoke freezing episodes during the turning tasks [22], the participants were asked to turn by making small steps on the spot instead of using big steps or a pirouette. An example of the turning task including a freezing episode is shown in the *supplementary multimedia file*. For each turning task, participants kept alternatively turning for 2 minutes, and each task was repeated maximal five times.

### D. Freezing annotations

Freezing episodes were annotated by two independent raters based on the videos taped during the tasks. Two video cameras: a GoPro Hero5 with a fisheye effect and a Sony HDR video camera were placed in the front and on the side of the participants, respectively. The two raters annotated freezing episodes and unexpected movements, such as sudden stops caused by other reasons than freezing episodes, based on the videos from the camera in the front. The videos from the camera on the side were used to assist the raters on uncertain freezing episodes.

The raters used an open-source annotation software (ANVIL [23]) for the annotations and had a substantial interrater agreement on the freezing annotations with 92% in percent-agreement and 0.78 in Cohen's kappa. A third rater made the final decision when the two raters disagreed on some annotations. Consequently, 551 freezing episodes during turning were agreed by the raters. The duration of the freezing episodes was from around one second to two minutes with a median duration of 3 s.

### E. Multi-modal signal collection

Multi-modal physiological signals of the participants were acquired during the tasks using an EEG cap (actiCAP, Brain Products GmbH) and a 32-channel portable system (Porti, Twente Medical Systems International B.V.). The EEG cap collected 60-channel EEG signals for brain activity and 4-channel electrooculography (EOG) signals for eye movements with a sampling rate of 500 Hz. The Porti system collected ECG signals for heart activity, 3D accelerometer signals above knees and ankles for motion activity, and footswitch signals for gait activity with a sampling rate of 512 Hz. For further analysis of signals, the ECG, accelerometers, and footswitches signals collected from the Porti system were down-sampled to 500 Hz in alignment with the EEG and EOG signals. The placement of the above mentioned sensors are presented in Fig. 1.

### F. Physiological features in multiple modalities

Physiological features were extracted from the signals of each modality. An overview of the multi-modal features is shown in Fig. 2.

(1) Brain activity (EEG signals)

Given the abnormal patterns in theta-band oscillations during the evolution of freezing episodes reported in the previous study [6], we calculated the power spectra within the theta band (4-7 Hz) from the EEG signals of the frontal-central channel (a subtraction between the signals of two electrodes: Fz and Cz). The raw signals were preprocessed using a low-pass filter (0.5-45 Hz). The power spectra were calculated using wavelet transformation with the Morlet wavelet [24].

(2) Eye movements (EOG signals)

During turning, quick and slow phases of the eye movements were observed consecutively, which represent the gaze shift for turning guidance and gaze stabilization (vestibulo ocular reflex), respectively. In this study, we analyzed the slow-phase velocity which indicates the speed for gaze resetting during turning and is defined as the first derivate of the horizontal EOG signals during the slow phases. The raw EOG signals were smoothed through a low-pass filter with a cutoff frequency of 30 Hz and a median filter with a window of 50 ms. In addition, the baseline drift of the signals, which is reconstructed from the coefficients of wavelet decomposition at the $10^{th}$ level, was removed to minimize noise [25]. The slow phases were detected using a well-performed K-means clustering algorithm, Cluster Fix [26].

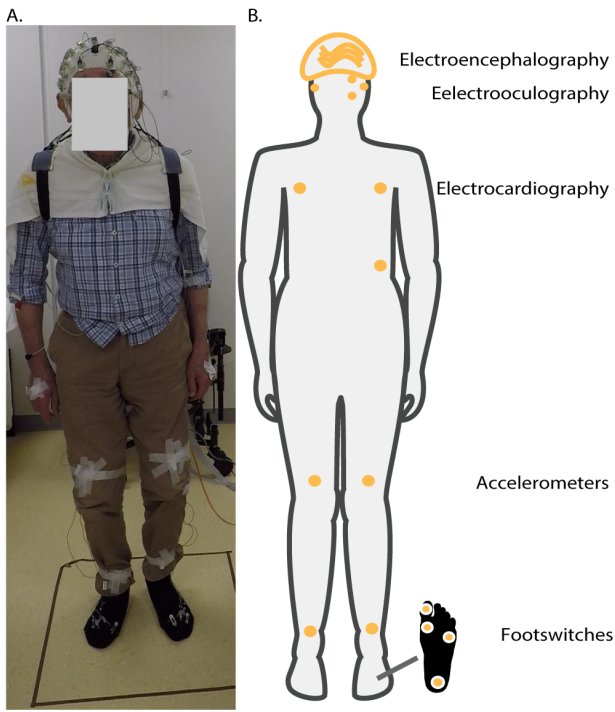

Fig. 1. Multi-modal sensor placement: (A) An example of a participant during the experiment. A video example about how participants executed the turning task and experienced freezing episode is presented in the *supplementary multimedia file*. (B) The illustration of the sensor placement whose signals were analyzed in this study.

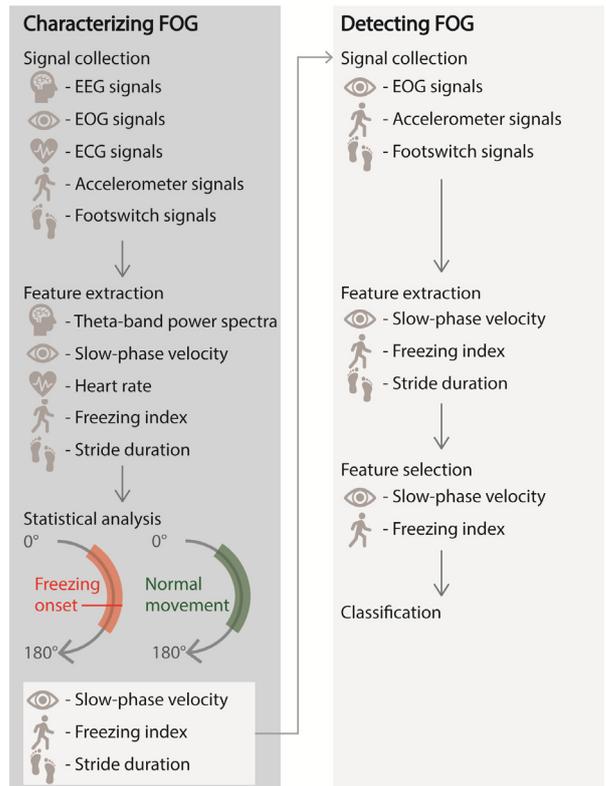

Fig. 2. An overview of the multi-modal features and the simplified workflow of the study: characterizing and detecting freezing of gait (FOG).

(3) Heart activity (ECG signals)

Heart rate for the heart activity was estimated from the 3-lead ECG signals. The baseline drift of the ECG signals was removed through the wavelet decomposition [25], and the R-peaks were detected by the widely used Pan-Tompkins algorithm [27]. The number of R-peaks within 1 minute is the heart rate (unit: bpm). To suppress movement artefacts and the noise of ECG signals, we calculated the heart rate measures of each ECG channels and treated the median heart rate as the value of heart rate per minute.

(4) Motion activity (accelerometer signals)

The values of freezing index were calculated from the 3D signals of accelerometers placed above knees and ankles to describe the motion activity of the lower body. Freezing index describes individual's trembling amplitude during movements and is defined as the ratio between the absolute power within 3–8 Hz for trembling and within 0.5–3 Hz for locomotion [5], [9]. The power spectrum was calculated using the wavelet transformation with the Morlet wavelet [24]. A sliding window of 2 s was used to calculate the freeze index at each sampling point (the window centered at the sampling point).

(5) Gait (footswitch signals)

Stride duration is a parameter to estimate the duration of a complete gait cycle (a cycle of both feet leaving and then striking the floor). In this study, we studied the gait cycle during turning. The status of each foot (leaving or striking the floor) was decided according to which of the footswitches on the ball of the foot was active (i.e., which parts of the foot touched the floor). The footswitch activation was determined through the comparison between the footswitch signals and a predefined voltage threshold matrix (offered by Twente Medical Systems International B.V.). Given the different foot-contact-floor patterns of individual participants, we define different set of switches for each individual participant to determine whether the foot contacted the floor. The setting of the key switches for individual participants was validated through visually checking their foot status in the taped videos.

### G. Statistical analysis

We used a two-sided t-test with a significance level of 0.05 to determine whether the means of two groups—physiological features before and during freezing episodes, and physiological features during normal movements (successful turning)—are significantly different, in other words, whether physiological features are significantly associated with freezing episodes. The multimodal physiological features were normalized across episodes to adjust the feature values on the different scales of participants and episodes.

Features from the $10^{th}$ s before the freezing onsets to the $3^{rd}$ s after the freezing onsets were analyzed because the median duration of the 551 freezing episodes was 3 s in our study, and we assumed that the value of a feature preceding a freezing episode with more than 10 s was not affected by the upcoming freezing episode. The corresponding feature segmentations in the normal-movement group were identified according to the phase angles of turning at the $10^{th}$ s before the freezing onsets. The phase angles of turning were calculated using the Hilbert transformation [28] from the slow-phase velocity indicating gaze stabilization during turning. Fig. 2 presents the illustration of the segmentations in the freezing (the red band in Fig. 2) and normal-turning (the green band in Fig. 2) group. In addition, only one freezing episode was included in the segmentations of the freezing group to suppress the influences of other freezing episodes, and no freezing episode was included in the corresponding segmentations of normal-movement group to minimize the effect of freezing episodes.

Given that we explored the changes of physiological features along the evolution of freezing episodes, we visualized the t-test through showing the mean and the 95% confidence intervals (mean ± standard error X 1.96) at each sampling point of the segmentations in the two groups. When the 95% confidence intervals of a physiological feature in the two groups does not overlap at a sampling point, the physiological feature is significantly associated with freezing episodes at the sampling point with the p-value < 0.05.

### H. Freezing of Gait Detection

The physiological features significantly associated with freezing episodes were used for the development of the freezing detection system. To prevent overfitting in the classification and improve the classification performances, we further ranked these features through the fast correlation-based feature selection method [29], [30]. In the feature selection method, symmetrical uncertainty, calculated as the normalized information gain between two variables and ranging from 0 to 1, estimates the correlations between features, and between features and classes [29]. A feature was selected when the symmetrical uncertainty between the feature itself and the classes was greater than 0.85 and the symmetrical uncertainty between the feature itself and other features.

The multi-modal features were extracted through a 50%-overlapping sliding window of 2.56 s from the pre-processed signals. The length of epochs, 2.56 s, was set given the requirement of an efficient-computing hardware architecture for a long-term real-time freezing detection system. Epochs were classified using a RUSBoost classifier, which is a commonly used classifier for solving an imbalance-class problem [31]. A freezing episode was detected (a true positive event) or not (a false negative event) based on whether any epoch was classified as the class of freezing episodes within the period from 3 seconds before the freezing onset to the freezing offset. The 3 seconds before the onsets was set as a buffer duration for strict freezing annotations. In addition, we treated each correctly and wrongly classified epoch in the class of normal movements as a true negative and false positive event, respectively. The detection system was evaluated using the leave-one-subject-out cross-validation method for the 15 participants. The performance of the freezing detection was estimated by its sensitivity, specificity, precision, and Matthews correlation coefficient (MCC) which is a correlation between true events and detected events, and a more reliable performance metric than accuracy and F1-score on imbalanced datasets [32].

We compared the performances of multi-modal (using features in multiple modalities) and single-modal (using features in only one modality) freezing detection systems through estimating Precision-Recall curves [33] and analyzing the values of MCC using a Wilcoxon signed-rank test [34]. Precision-Recall curves can avoid the over-optimism of Receiver Operator Characteristic curves in the severely imbalanced freezing classification problem [33]. The different values of the recall (sensitivity) and their corresponding precision values were generated through changing the threshold of the scores of each epoch in the RUSBoost classifier. Moreover, the discontinuous precision values were interpolated [33], and the area under the curves was approximated using the composite trapezoidal method. In this study, we used the Wilcoxon signed-rank test, a non-parametric statistical test as an alternative to paired t-test [34], to check whether the median difference of MCC between the multi-modal system and individual single-modal systems is zero. Given the multiple comparisons, Bonferroni correction was used to control the familywise error rate [35]. When the p-value is smaller than 0.05, the performance difference between the multi-modal and single-modal systems is statistically significant.

## III. RESULTS

In total, 551 freezing episodes (ranged 1 s-2 min with a median of 3 s) in turning tasks were annotated from 15 participants and used for the training and testing of the freezing detection systems. 248 of the 551 freezing episodes, which did not have neighboring freezing episodes in the segmentations of freezing-group, were investigated for the statistical analysis. In addition, one subject's ECG signals were not applicable because of a bad connection between the ECG sensors and the skin during the experiment, therefore 30 freezing episodes of this subject were not included for the statistical analysis of heart rate.

The visualization of the t-test of the segmentations in the freezing and normal-movement groups are presented in Fig. 3-7. Significant changes associated with freezing episodes were observed in the motion activity described by the freezing index above the knees and ankles (Fig. 3), the eye movements described by the slow-phase velocity (Fig. 4), and the gait described by the stride duration (Fig. 5). The brain activity described by the power spectra of frontal-central channel (Fig. 6) and the heart activity described by the heart rate (Fig. 7) were not closely associated with freezing episodes.

The development of the freezing detection system was based on the results of the statistical analysis. Accordingly, the slow-phase velocity, freezing index, and stride duration were extracted from the epochs of the preprocessed EOG, accelerometer, and footswitch signals, respectively. Through the correlation-based feature selection, the slow phase velocity and freezing index above the left knee, whose symmetrical uncertainty were greater than 0.85, were selected for the classification.

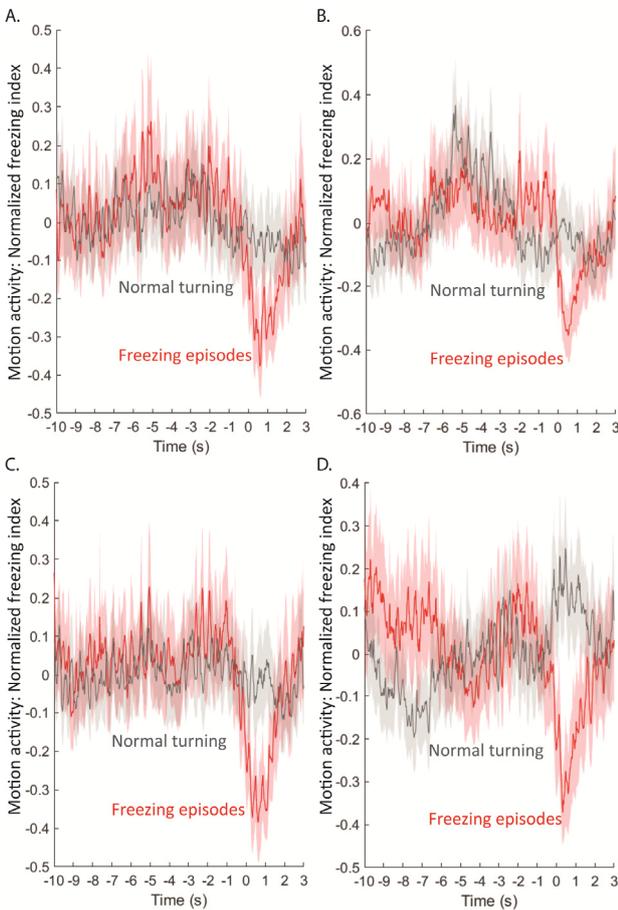

Fig. 3. Motion activity: a decrease of the normalized freezing index during freezing episodes. The freezing index was calculated from the accelerometers placed above knees [left: (A); right (B)] and ankles [left: (C); right (D)]. The red curve indicates the freezing index in the group of freezing episodes, and the gray curve indicates the freezing index during normal turning. The solid lines indicate the mean of the freezing index, and the patches indicate 95% confidence intervals across segmentations.

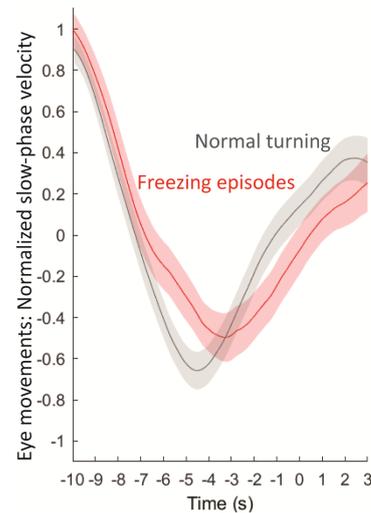

Fig. 4. Eye movements: the normalized slow-phase velocity decreased around 6 seconds before the onsets of freezing episodes. The red curve indicates the group of freezing episodes, and the gray curve indicates the group of normal turning. The solid lines indicate the mean of the slow-phase velocity, and the patches indicate 95% confidence intervals across segmentations. The values in the clockwise turns and the counterclockwise turns were merged through flipping the values of the slow-phase velocity during counterclockwise turns.

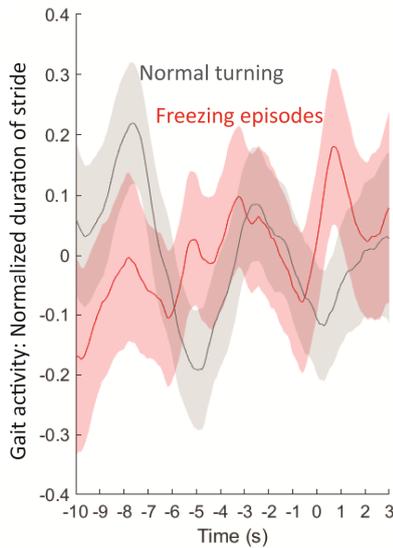

Fig. 5. Gait activity: the normalized duration of stride (a complete gait cycle) estimated from footswitch signals. The stride duration shows an considerable increase during freezing episodes. The red curve indicates the stride duration in the group of freezing episodes, and the gray curve indicates the stride duration during normal turning. The solid lines indicate the mean of the stride duration, and the patches indicate 95% confidence intervals across segmentations. The increases of the curves between -10 seconds and -8 second, and between the -5 and -3 second was probably caused by short stops between the normal turns.

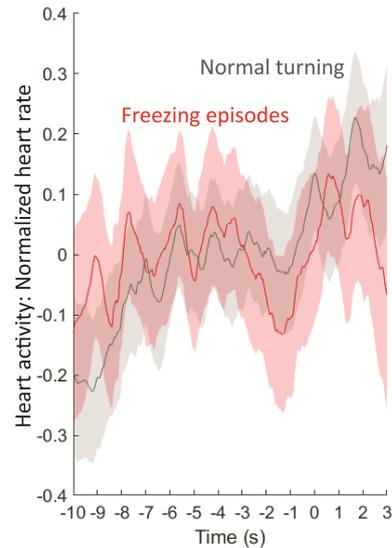

Fig. 7. Heart activity: normalized heart rate estimated from electrocardiography signals: No considerable difference of heart rate was observed the groups of freezing episodes (red curves) and normal turning (gray curves). The drops of both curves between the -3 and -1 second were probably caused by the stops between turns. The solid lines indicate the mean of heart rate, and the patches indicate 95% confidence intervals across segmentations.

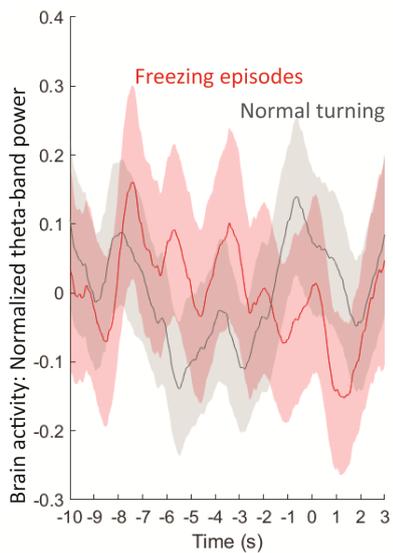

Fig. 6. Brain activity: No considerable difference of normalized theta-band power in the frontal-central electroencephalography channel was observed between the groups of freezing episodes (red curves) and normal turning (gray curves). The solid lines indicate the mean of theta-band power, and the patches indicate 95% confidence intervals across segmentations.

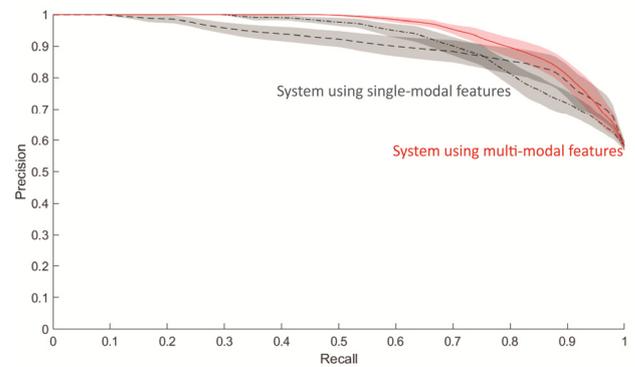

Fig. 8. Precision-Recall curves of the freezing detection systems using multi-modal (red curve) and single-modal (gray curves) features. The multi-modal features merged the features from the motion activity (dashed line) described by freezing index and the eye-movements (dash-dotted line) described by slow-phase velocity. Generally, the freezing detection system using multi-modal features performed better than using the single-modal features. The solid lines indicate the mean value across participants, and the patches indicate the standard error.

The performance of the multi-modal freezing detection system with the freezing index and the slow phase velocity was 97% ± 3% of sensitivity, 96% ± 7% of specificity, 73% ± 21% of precision, and 0.82 ± 0.15 of MCC. The comparison through Precision-Recall curves between the multi-modal system and the corresponding single-modal systems, which only used the freezing index above left knee or the slow phase velocity, is shown in Fig. 8. The areas under the Precision-Recall curves of the freezing detection systems using multi-modal (solid line) and single-modal features [using the freezing index (dash line) or the slow phase velocity (dash-dotted line)], were 0.94 ± 0.05, 0.91 ± 0.08, and 0.90 ± 0.08, respectively. Nevertheless, we did not find significant differences of MCC between the multi-modal system and the single-modal systems through the Wilcoxon signed-rank test.

## IV. Discussion

To improve the reliability of freezing detection and fill in the knowledge gap of online freezing detection systems using multi-modal features, we characterized freezing episodes provoked by turns using the parameter of eye movements, brain, heart, motion, and gait activity, and proposed an online freezing detection system using the top-ranked features significantly associated with freezing episodes. The parameter of eye movements, gait activity, and motion activity were found to be significantly associated with freezing episodes. In contrast to earlier findings [6], [7], heart activity described by the heart rate and brain activity described by the power spectra of the frontal-central channel were not associated with freezing episodes. Through the feature selection, the freezing index above left knee and the slow phase velocity were selected as the features in the detection of freezing episodes. The detection system's sensitivity was of 97% ± 3%, specificity was of 96% ± 7%, precision was of 73% ± 21%, and MCC was of 0.82 ± 0.15. Through the comparison using Precision-Recall curves, the detection system using multi-modal features performed better than the single-modal features, which further supported the idea in our previous study [14]. However, the results of the Wilcoxon signed-rank test showed no significant differences between the performance of multi-modal system and single-modal system.

The abnormal patterns found in the slow-phase velocity indicates that the participants' eye-resetting speed during turning has slowed down almost 6 seconds before freezing onsets. The considerable changes of the slow-phase velocity can be explained from two possible aspects: (1) The excessive inhibition of the superior colliculus projected from the basal ganglia via the substantia nigra pars reticulata may have an effect on ocular motor engagement [36]; (2) The slowness or even stop of turning could have an influence on the eye movements. In this study, the angles of the body were not measured to investigate the slowness of body during turning. Future research should be undertaken to investigate the relationship between the eye, head, and body movements under turning conditions surrounding freezing episodes. In addition, turning-phase angles may have influences on the slow-phase velocity given the vestibulo ocular reflex during turning. In our future work, the slow-phase velocity should be divided into several groups according to the turning-phase angles for the deeper investigation of the eye movements.

The significant changes of freezing index and stride duration during freezing episodes consist with the definition of freezing-of-gait that individuals' lower body barely moves during the freezing episodes. In our study, the freezing index considerably decreased during freezing episodes instead of increasing reported in the previous study [5]. The decrease of the power in the freezing frequency band (3-8 Hz) in our study may be explained by the influence of the guidance on continuous turning conditions to provoke enough freezing episodes (the participants were asked to use the full surface of their feet while turning and turn by making small steps on the spot; in other words, not to use big steps or a pirouette). Through the visual analysis of the videos, we found that participants showed less trembling symptoms and almost stopped turning or made slow small steps when they experienced freezing episodes (an example is shown in the *supplementary multimedia file*). Future studies should investigate the influence of different turning guidance on the values of freezing index and the freezing detection. In addition, the freezing index above left knee was top-ranked through the correlation-based feature selection method for freezing detection. We speculated that the contrast above left knee between movement changes caused by freezing and normal movements may be stronger than other lower body parts. This might be caused by the asymmetric motor function in individuals with Parkinson's disease [37]. Following studies should consider the asymmetry of the individuals in the freezing detection.

In this study, we first analyzed which physiological features in multiple modalities are associated with freezing episodes to globally characterize freezing episodes for the evidence-support of freezing detection and duplicate the results reported in previous studies [6], [7] where movement artefacts were not considered. We statistically analyzed the physiological features in the freezing and normal-turning groups to consider the potential disturbances of the turning conditions on the freezing analysis. This probably explains why our results did not support the earlier findings about the heart activity [7] and brain activity [6]. In those previous research, the parameters of the heart and brain activity were compared among the freezing stages (the periods before, during, and after freezing episodes). Freezing episodes were relatively sensitive to the movement conditions. For example, turning is the most sensitive condition to provoke freezing episodes [21], [22]. The previous studies most likely compared the parameters between movement status, such as between walking and turning, instead of between the stages of freezing episodes. For example, an increase of heart rate was observed in both groups of freezing episodes and normal turning between the -3$^{rd}$ and 1$^{st}$ s in Fig. 7. In the future temporal analysis of freezing episodes, we strongly suggest a comparison of parameters in a freezing group with the corresponding baseline movement condition to suppress the influence of movement. In the future studies about the freezing detection or prediction, we recommend to use the evidence-support features in the development to improve the reliability of systems.

The multi-modal freezing detection system performed better than the single-modal systems according to the Precision-Recall curves, but the difference is not significant through the Wilcoxon signed-rank test on 15 participants. The contrast between the results of the Precision-Recall curves and the Wilcoxon signed-rank test may be explained by the small number of participants which limited the power of the Wilcoxon signed-rank test to find a statistical significance. In future work, the power of the statistical test could be increased with a larger number of participants. In addition, the performance from the multi-modal and single-modal systems probably do not reflect the performance under real-life situations given that the dataset of this study was acquired during continuous turning conditions in a lab environment. Future studies are needed to investigate the system performances under a (simulated) real-life situations to further determine whether the multi-modal system performs better than using single-modal features. Moreover, practically, users may prefer the single-modal systems because of its simple application in daily life. In future studies, it is valuable to investigate whether it is acceptable for users to have multiple types of sensors on their body for daily usage.

Several limitations still remain in this study: only 17 participants were included in this explorative study, and the physiological signals were acquired from 15 participants while the participants were at off-medication state. The generalization of the results in this study may be influenced by the small number of participants, and an effect of the dopaminergic treatment on physiological signals might exist. More participants at on-medication state (with the effect of the dopaminergic treatment) should be included in future investigations to ensure the transformation from research to practical applications. Moreover, only one commonly used feature of each modality was investigated in this study. In future studies, more features for each modality should be investigated, such as quick-phase amplitude indicating gaze shift during turning for eye movements, heart variability for heart activity, galvanic skin response for anxiety level, brain connectivity for brain activity, and swing and stance duration for gait. Given that the all participants of this study were not treated by deep brain stimulus, our proposed freezing detection system for the population with deep brain stimulus should be further validated and expanded in future. For example, the power in high-beta band (21-35 Hz), whose freezing-associated changes is found in subthalamic nucleus using intracranial EEG signals in other earlier study [38], should be considered as a feature for freezing detection in the population with deep brain stimulation. Furthermore, different fusion methods for multi-modal features, such as principle component analysis, would be interesting to be investigated in the freezing detection. In addition, this study focused on the freezing detection (from 3 seconds before freezing onsets to freezing offsets). Future research should be undertaken to investigate how early and accurate we can predict freezing episodes.

## V. Conclusion

Multi-modal physiological signals were used to characterize and detect freezing episodes in this study. The slow-phase velocity for eye movements and freezing index for motion activity were significantly associated with freezing episodes and selected as multi-modal features for freezing detection. Using the multi-modal features, the freezing detection performance was improved according to Precision-Recall curves. Future works need to further validate and improve the detection system using multi-modal features in a real-life situation.

## Appendix

A demo video clip of the turning task from a participant is presented in supplementary multimedia file

## Acknowledgment

We are grateful to all participants for their time and effort. This work was strongly supported by Dutch Parkinson's disease Association (Parkinson Vereniging). We especially thanks Jan Gouman, Piet van de Schilde, and Erik Jan Marinissen for their precious suggestions. In addition, we are also grateful to the contribution during the experiments and data analysis from the assistants: Sabine Janssen, Cat de Win, Günter Windau, Jessica Askamp, Jamie Jansen, Karlijn van Dijsseldonk, Nynke Tilkema, Rowena Emaus, and Rowena van der Velden.

## References

[1] J. G. Nutt, B. R. Bloem, N. Giladi, M. Hallett, F. B. Horak, and A. Nieuwboer, "Freezing of gait: moving forward on a mysterious clinical phenomenon," *Lancet Neurol.*, vol. 10, no. 8, pp. 734–744, Aug. 2011.

[2] B. R. Bloem, J. M. Hausdorff, J. E. Visser, and N. Giladi, "Falls and freezing of gait in Parkinson's disease: A review of two interconnected, episodic phenomena," *Mov. Disord.*, vol. 19, no. 8, pp. 871–884, Aug. 2004.

[3] M. Gilat, A. Lígia Silva de Lima, B. R. Bloem, J. M. Shine, J. Nonnekes, and S. J. G. Lewis, "Freezing of gait: Promising avenues for future treatment," *Park. Relat. Disord.*, vol. 52, pp. 7–16, Jul. 2018.

[4] N. Giladi and A. Nieuwboer, "Understanding and treating freezing of gait in parkinsonism, proposed working definition, and setting the stage," *Mov. Disord.*, vol. 23, no. S2, pp. S423–S425, Jul. 2008.

[5] S. T. Moore, H. G. MacDougall, and W. G. Ondo, "Ambulatory monitoring of freezing of gait in Parkinson's disease," *J. Neurosci. Methods*, vol. 167, no. 2, pp. 340–348, Jan. 2008.

[6] J. M. Shine *et al.*, "Abnormal patterns of theta frequency oscillations during the temporal evolution of freezing of gait in Parkinson's disease," *Clin. Neurophysiol.*, vol. 125, no. 3, pp. 569–576, Mar. 2014.

[7] I. Maidan, M. Plotnik, A. Mirelman, A. Weiss, N. Giladi, and J. M. Hausdorff, "Heart rate changes during freezing of gait in patients with Parkinson's disease," *Mov. Disord.*, vol. 25, no. 14, pp. 2346–2354, Oct. 2010.

[8] S. Mazilu, A. Calatroni, E. Gazit, A. Mirelman, J. M. Hausdorff, and G. Tröster, "Prediction of freezing of gait in Parkinson's from physiological wearables: An exploratory study," *IEEE J. Biomed. Heal. Informatics*, vol. 19, no. 6, pp. 1843–1854, Nov. 2015.

[9] M. Bachlin *et al.*, "Wearable Assistant for Parkinson's Disease Patients With the Freezing of Gait Symptom," *IEEE Trans. Inf. Technol. Biomed.*, vol. 14, no. 2, pp. 436–446, Mar. 2010.

[10] A. M. A. Handojoseno, J. M. Shine, T. N. Nguyen, Y. Tran, S. J. G. Lewis, and H. T. Nguyen, "The detection of Freezing of Gait in Parkinson's disease patients using EEG signals based on Wavelet decomposition," in *2012 Annual International Conference of the IEEE Engineering in Medicine and Biology Society*, 2012, pp. 69–72.

[11] A. M. A. Handojoseno, J. M. Shine, T. N. Nguyen, Y. Tran, S. J. G.

Lewis, and H. T. Nguyen, "Using EEG spatial correlation, cross frequency energy, and wavelet coefficients for the prediction of Freezing of Gait in Parkinson's Disease patients," in *2013 35th Annual International Conference of the IEEE Engineering in Medicine and Biology Society (EMBC)*, 2013, pp. 4263–4266.

[12] J. D. Schaafsma, Y. Balash, T. Gurevich, A. L. Bartels, J. M. Hausdorff, and N. Giladi, "Characterization of freezing of gait subtypes and the response of each to levodopa in Parkinson's disease," *Eur. J. Neurol.*, vol. 10, no. 4, pp. 391–398, Jul. 2003.

[13] Pardoel, Kofman, Nantel, and Lemaire, "Wearable-Sensor-based Detection and Prediction of Freezing of Gait in Parkinson's Disease: A Review," *Sensors*, vol. 19, no. 23, p. 5141, Nov. 2019.

[14] Y. Wang *et al.*, "Freezing of gait detection in Parkinson's disease via multimodal analysis of EEG and accelerometer signals," in *Engineering in Medicine and Biology Society (EMBC) 2020 Annual International Conference of the IEEE*, 2020.

[15] C. G. Goetz *et al.*, "Movement Disorder Society-sponsored revision of the Unified Parkinson's Disease Rating Scale (MDS-UPDRS): Scale presentation and clinimetric testing results," *Mov. Disord.*, vol. 23, no. 15, pp. 2129–2170, Nov. 2008.

[16] M. M. Hoehn and M. D. Yahr, "Parkinsonism: onset, progression, and mortality," *Neurology*, vol. 17, no. 5, pp. 427–427, May 1967.

[17] B. K. Scanlon, H. L. Katzen, B. E. Levin, C. Singer, and S. Papapetropoulos, "A formula for the conversion of UPDRS-III scores to Hoehn and Yahr stage," *Parkinsonism Relat. Disord.*, vol. 14, no. 4, pp. 379–380, May 2008.

[18] A. Nieuwboer *et al.*, "Reliability of the new freezing of gait questionnaire: Agreement between patients with Parkinson's disease and their carers," *Gait Posture*, vol. 30, no. 4, pp. 459–463, Nov. 2009.

[19] M. F. Folstein, "The Mini-Mental State Examination," *Arch Gen Psychiatry*, vol. 40, no. 7, p. 812, Jul. 1983.

[20] B. Dubois, A. Slachevsky, I. Litvan, and B. Pillon, "The FAB: A frontal assessment battery at bedside," *Neurology*, vol. 55, no. 11, pp. 1621–1626, Dec. 2000.

[21] K. van Dijsseldonk, Y. Wang, R. van Wezel, B. R. Bloem, and J. Nonnekes, "Provoking Freezing of Gait in Clinical Practice: Turning in Place is More Effective than Stepping in Place," *J. Parkinsons. Dis.*, vol. 8, no. 2, pp. 363–365, Jun. 2018.

[22] J. Nonnekes, A. M. Janssen, S. H. G. Mensink, L. B. Oude Nijhuis, B. R. Bloem, and A. H. Snijders, "Short rapid steps to provoke freezing of gait in Parkinson's disease," *J. Neurol.*, vol. 261, no. 9, pp. 1763–1767, Sep. 2014.

[23] M. Kipp, "Multimedia Annotation, Querying, and Analysis in Anvil," in *Multimedia Information Extraction*, Hoboken, NJ, USA: John Wiley & Sons, Inc., 2012, pp. 351–367.

[24] M. X. Cohen, "Frequency and Time-Frequency Domains Analyses," in *Analyzing Neural Time Series Data: Theory and Practice*, MIT Press, 2014, pp. 109–272.

[25] A. Bulling, J. A. Ward, H. Gellersen, and G. Tröster, "Eye Movement Analysis for Activity Recognition Using Electrooculography," *IEEE Trans. Pattern Anal. Mach. Intell.*, vol. 33, no. 4, pp. 741–753, Apr. 2011.

[26] S. D. König and E. A. Buffalo, "A nonparametric method for detecting fixations and saccades using cluster analysis: Removing the need for arbitrary thresholds," *J. Neurosci. Methods*, vol. 227, pp. 121–131, Apr. 2014.

[27] J. Pan and W. J. Tompkins, "A Real-Time QRS Detection Algorithm," *IEEE Trans. Biomed. Eng.*, vol. BME-32, no. 3, pp. 230–236, Mar. 1985.

[28] S. Lawrence Marple, "Computing the discrete-time analytic signal via fft," *IEEE Trans. Signal Process.*, vol. 47, no. 9, pp. 2600–2603, Sep. 1999.

[29] M. A. Hall, "Correlation-based Feature Selection for Discrete and Numeric Class Machine Learning," in *Proceedings of the Seventeenth International Conference on Machine Learning*, 2000, pp. 359–366.

[30] L. Yu and H. Liu, "Feature Selection for High-dimensional Data: A Fast Correlation-based Filter Solution," in *Proceedings of the Twentieth International Conference on International Conference on Machine Learning*, 2003, pp. 856–863.

[31] C. Seiffert, T. M. Khoshgoftaar, J. Van Hulse, and A. Napolitano, "RUSBoost: A Hybrid Approach to Alleviating Class Imbalance," *IEEE Trans. Syst. Man, Cybern. - Part A Syst. Humans*, vol. 40, no. 1, pp. 185–197, Jan. 2010.

[32] D. Chicco and G. Jurman, "The advantages of the Matthews correlation coefficient (MCC) over F1 score and accuracy in binary classification evaluation," *BMC Genomics*, vol. 21, no. 1, p. 6, Dec. 2020.

[33] J. Davis and M. Goadrich, "The Relationship Between Precision-Recall and ROC Curves," in *Proceedings of the 23 rd International Conference on Machine Learning*, 2006, pp. 233–240.

[34] J. H. McDonald, "Wilcoxon signed-rank test," in *Handbook of Biological Statistics, 3rd edition*, Baltimore, Maryland: Sparky House Publishing, 2014, pp. 187–190.

[35] J. H. McDonald, "Multiple comparisons," in *Handbook of Biological Statistics, 3rd edition*, Baltimore, Maryland, 2014, pp. 257–263.

[36] Y. Terao *et al.*, "Initiation and inhibitory control of saccades with the progression of Parkinson's disease - Changes in three major drives converging on the superior colliculus," *Neuropsychologia*, vol. 49, no. 7, pp. 1794–1806, Jun. 2011.

[37] M. Plotnik, N. Giladi, Y. Balash, C. Peretz, and J. M. Hausdorff, "Is freezing of gait in Parkinson's disease related to asymmetric motor function?," *Ann. Neurol.*, vol. 57, no. 5, pp. 656–663, May 2005.

[38] J. B. Toledo *et al.*, "High beta activity in the subthalamic nucleus and freezing of gait in Parkinson's disease," *Neurobiol. Dis.*, vol. 64, pp. 60–65, Apr. 2014.